\documentclass[fleqn]{annalen}

\begin{document}

%%%%%%%% the following newcommands will be completed by the publisher %%%%%%%%

\newcommand{\volume}{9}              %sets current volume,
\newcommand{\xyear}{2000}            %sets year in header
\newcommand{\issue}{5}               %sets current issue,
\newcommand{\recdate}{29 Aug. 2000}  %sets received date,
\newcommand{\revdate}{dd.mm.yyyy}    %sets revised date,
\newcommand{\revnum}{0}              %number of revisions,
\newcommand{\accdate}{25 Sep. 2000}    %sets accepted date,
\newcommand{\coeditor}{C. Thomsen}   %sets (co)editor,
\newcommand{\firstpage}{855}         %first page number,
\newcommand{\lastpage}{864}          %last page number,
\setcounter{page}{\firstpage}        %sets page counter to first page number
%%%%%%%%%%%%%%%%%% please give up to three keywords here %%%%%%%%%%%%%%%%%%%%%
\newcommand{\keywords}{Decoherence, Einselection, Information}
%%%%%%%%%%%%%%%% please give up to three PACS numbers here %%%%%%%%%%%%%%%%%%%
\newcommand{\PACS}{03.65.Bz, 89.70}
%% please enter (First) Author (et al.) and short version of the title here %%
%%%%%%%%%%%% must not exceed 80 characters in length together %%%%%%%%%%%%%%%%
\newcommand{\shorttitle}{W. H. Zurek, Einselection and Decoherence}
%% sets the header on oddpage
%%%%%%%%%%%%%%%%%%%%%%%% here comes the title group %%%%%%%%%%%%%%%%%%%%%%%%%%
\title{Einselection and Decoherence \\from an Information Theory
Perspective}

%%%%%%%%%%%%%%%%%%%%%%%%%%%%%%%%%%%%%%%%%%%%%%%%%%%%%%%%%%%%%%%%%%%%%%%%%%%%%%

\author{W. H. Zurek}

%%%%%%%%%%%%%%%%%%%%%%%%%%%%%%%%%%%%%%%%%%%%%%%%%%%%%%%%%%%%%%%%%%%%%%%%%%%%%%

\newcommand{\address}{Los Alamos National Lab. \\ T-6, Los Alamos, NM 87545
\hspace*{0.5mm} USA}

%%%%%%%%%%%%%%%%%%%%%%%%%%%%%%%%%%%%%%%%%%%%%%%%%%%%%%%%%%%%%%%%%%%%%%%%%%%%%%

\newcommand{\email}{\tt whz@lanl.gov}

\maketitle

%%%%%%%%%%%%%%%%%%%%%%%%%%%%%%%%%%%%%%%%%%%%%%%%%%%%%%%%%%%%%%%%%%%%%%%%%%%%%

\begin{abstract}

We introduce and investigate a simple model of conditional quantum
dynamics. It allows for a discussion of the information-theoretic
aspects of quantum measurements, decoherence, and
environment-induced superselection (einselection).

\end{abstract}

%%%%%%%%%%%%%%%%%%%%%%%%%%%%%%%%%%%%%%%%%%%%%%%%%%%%%%%%%%%%%%%%%%%%%%%%%%%%%

\section{Introduction}

Transfer of information was the focus of attention [1-3] of
research on decoherence since the early days. In the intervening
two decades this perspective was not forgotten [4], but the study
of different mechanisms of decoherence [5-9] took precedence over
considerations of information-theoretic nature. The aim of this
paper is to sketch a few ideas which tie the ``traditional''
points of view of einselection and decoherence (especially the
issue of the preferred pointer basis) to various other aspects of
decoherence that have a strong connection with
information-theoretic concepts.

A large part of our discussion shall be based on a simple model of
conditional dynamics, which is a direct generalization of the
``bit by bit'' measurement introduced in [1] and studied in [3].
We shall introduce the model in Section 2 and use it to compute
the ``price" of information in units of action in Section 3.
Section 4 defines information theoretic quantum {\it discord}
between two classically identical definitions of mutual
information. Discord can be regarded as a measure of a violation
of classicality of a joint state of two quantum subsystems.
Section 5 turns to the evolution of the state of the environment
in course of decoherence. The {\it redundancy ratio} introduced
there can be regarded as a measure of objectivity of quantum
states. A large redundancy ratio is a sufficient condition for an
effective classicality of quantum states.

\section{Controlled shifts for conditional dynamics}

The simplest example of an entangling quantum evolution is known
as the controlled not ({\tt c-not}). It involves two bits (a
``control'' and a ``target''). Their interaction leads to:
\newcounter{saveeqn}
\newcommand{\alpheqn}{\setcounter{saveeqn}{\value{equation}}%
\setcounter{equation}{0}%
\renewcommand{\theequation}{%
\mbox{\arabic{saveeqn}\alph{equation}}}}%
\newcommand{\reseteqn}{\setcounter{equation}{\value{saveeqn}}%
\renewcommand{\theequation}{\arabic{equation}}}
\alpheqn \setcounter{saveeqn}{1}
\begin{equation}
  |0_C\rangle |x_T\rangle  \longrightarrow   |0_C\rangle|x_T\rangle
\end{equation}
\begin{equation}
|1_C\rangle |x_T\rangle   \longrightarrow  |1_C\rangle|\neg
x_T\rangle
\end{equation}
where the state $|\neg x\rangle$ is defined through a
basis-dependent negation: \reseteqn
\begin{equation}
\neg (\gamma|0_T\rangle+\eta|1_T\rangle) =\gamma| 1_T\rangle +
\eta| 0_T\rangle \ .
\end{equation}
Classical {\tt c-not} ``flips'' the target bit whenever the
control is in the state ``1'', but does nothing otherwise.
Quantum {\tt c-not} is an obvious generalization.

The distinction between the classical and quantum {\tt c-not}
comes from the fact that both quantum and classical bits can be
in an arbitrary superposition. Thus, {\tt c-not} starting from a
superposition of $|0\rangle$ and $|1\rangle$ will in general lead
to an entangled state. Moreover, when both the control and the
target start in Hadamard-transformed states:
\begin{equation}
|\pm\rangle = (|0\rangle \pm |1\rangle)/\sqrt{2} \ ,
\end{equation}
{\tt c-not} reverses direction: \alpheqn \setcounter{saveeqn}{4}
\begin{equation}
|\pm\rangle|+\rangle \longrightarrow |\pm\rangle|+\rangle ;
\end{equation}
\begin{equation}
|\pm\rangle|-\rangle \longrightarrow |\mp\rangle|-\rangle \ .
\end{equation}
\reseteqn \noindent Above, we have dropped labels: The original
control is always to the left, as was the case in Eq. (1). We say
``the original'', because the Hadamard transform of Eq. (3)
reverses the direction of the information flow in the quantum
{\tt c-not}. As can be seen in Eq. (4), the sign of the former
control (left ket) flips when the former target is in the state
$|-\rangle$.

Study of such simple models has led to
the concept of preferred pointer states [1] and einselection
[2,3]. We shall not review here these well known developments,
directing the reader instead to the already available [10] or
forthcoming [11] reviews of the subject.

Controlled shift ({\tt c-shift}) is a straightforward generalization
of {\tt c-not}. The original truth table (an analogue of Eq. (1)) can be
written as:
\begin{equation}
|s_j\rangle |A_k\rangle \longrightarrow |s_j\rangle
|A_{k+j}\rangle \
\end{equation}
There is also a control and a target (which we shall more often
call ``the system ${\cal S}$" and ``the apparatus ${\cal A}$",
reflecting this nomenclature in notation). Equation (5) implies
Eq. (1) when both ${\cal S}$ and ${\cal A}$ have two-dimensional
Hilbert spaces. Moreover, when $j=0$, Eq. (1) becomes a model of a
pre-measurement:
\begin{equation}
|\Psi_0\rangle = |\psi\rangle |A_0\rangle = \left ( \sum_i a_i
|s_i\rangle \right )|A_0\rangle
\longrightarrow
\sum_i a_i |s_i\rangle |A_i\rangle = |\Psi_t\rangle\ .
\end{equation}
As in the case of {\tt c-not}, in the respective bases $\{
|s_i\rangle \}$ and $\{|A_k\rangle\}$, Eq. (5) seems to imply a
one - way flow of information, from ${\cal S}$ to ${\cal A}$.
However, a complementary basis [12,13] can be readily defined:
\begin{equation}
|B_k\rangle\ =\ N^{ - {1 \over 2 }} \sum_{l=0}^{N-1}\exp ({{2 \pi
i} \over N} kl) ~ |A_l\rangle \ .
\end{equation}
It is analogous to the Hadamard transform we have introduced
before, but it also has an obvious affinity to the Fourier
transform. We shall call it a Hadamard-Fourier Transform (HFT).
It is straightforward to show that the orthonormality of
$\{|A_k\rangle\}$ immediately implies:
\begin{equation}
\langle B_l | B_m \rangle = \delta _{lm} \ .
\end{equation}
The inverse of HFT can be easily given:
\begin{equation}
|A_k\rangle\ =\ N^{ - {1 \over 2 }} \sum_{l=0}^{N-1}\exp (-{{2
\pi i} \over N} kl) ~ |B_l\rangle \ .
\end{equation}
Consequently, for an arbitrary $|\psi\rangle$;
\begin{equation}
|\psi\rangle = \sum_n \alpha_n |A_n\rangle = \sum_k \beta_k
|B_k\rangle \ ,
\end{equation}
where the coefficients are given by the HFT;
\begin{equation}
\beta_k = N^{-\frac{1}{2}} \sum_{n=0}^{N-1} \exp(-\frac{2 \pi i}
{N} kn) \alpha_n  .
\end{equation}
To implement the truth table of Eq. (5) we shall use observables
of the apparatus: \alpheqn \setcounter{saveeqn}{12}
\begin{equation}
\hat A = \sum_{k=0}^{N-1} k |A_k\rangle \langle A_k|;
\end{equation}
\begin{equation}
\hat B = \sum_{l=0}^{N-1} l |B_l\rangle\langle B_l| \ ,
\end{equation}
\reseteqn
as well as the observable of the system:
\begin{equation}
\hat s =  \sum_{l=0}^{N-1} l |s_l\rangle\langle s_l| \ .
\end{equation}
The interaction Hamiltonian of the form;
\begin{equation}
H_{int} = g \hat s \hat B \
\end{equation}
acting over a period $t$ will induce a transition;
\begin{equation}
\exp(-i H_{int} t / \hbar) |s_j\rangle |A_k\rangle = |s_j> N^{-{1
\over 2}} \sum_{l=0}^{N-1}\exp[-i(jgt/\hbar + 2\pi k /N)l]
|B_l\rangle \ .
\end{equation}
Thus, if the coupling constant $g$ is selected so that the
associated action is
\begin{equation}
I = gt/\hbar =  G \times 2\pi / N \ .
\end{equation}
a perfect one-to-one correlation between the states of the system
and of the apparatus can be accomplished, since:
\begin{equation}
\exp(-i H_{int} t / \hbar) |s_j\rangle |A_k\rangle\ = \
|s_j\rangle |A_{{\{k+G * j\}}_N}\rangle \ .
\end{equation}
Thus, true to its name, the interaction described here
accomplishes a simple shift, Eq. (5), of the state of the
apparatus, while the system acts as a control. The index $\{k +
G*j\}_N$ has to be evaluated modulo $N$ (where $N$ is the number
of the orthogonal states of the apparatus) so that when $k+G*j >
N$, the apparatus states can ``rotate'' through $|A_0\rangle$ and
stop where the interaction takes it. The integer $G$ can be
regarded as the gain factor. As Eqs. (14) - (17) imply, the
adjacent states of the system (i.e.,
$|s_j\rangle,~|s_{j+1}\rangle$) get mapped onto the states of the
apparatus that are $G$ apart ``on the dial''.

When the dimension of the Hilbert space of the system $n$ is such
that
\begin{equation}
n G < N  \ ,
\end{equation}
the above model provides one with a simple example of
amplification. It is possible to use it to study the utility of
amplification in increasing signal to noise ratio in measurements
[11]. It also shows why amplification can bring about decoherence
and effective irreversibility (although {\tt c-shift} is of
course perfectly reversible). We shall employ {\tt c-shift} to
study the cost of information transfer, to introduce information
-- theoretic discord, the measure of the classicality of correlations,
and to discuss objectivity of quantum states which arises from the
redundancy of the records imprinted by the state of the system on
its environment.

\section {Planck's constant and the price of a bit}

Transfer of information is the objective of the measurement
process and an inevitable consequence of most interactions. It
happens in course of decoherence. Here we shall quantify its cost
in the units of action.

The consequence of the interaction between ${\cal S}$ and ${\cal
A}$ is the correlated state $|\Psi_t\rangle$, Eq. (6). While the
joint state of ${\cal AS}$ is pure, each of the subsystems is in a
mixed state given by the reduced density matrix of the system
\alpheqn \setcounter{saveeqn}{19}
\begin{equation}
\rho_{\cal S} \ =  \mbox{Tr}_{\cal A}
|\Psi_t\rangle\langle\Psi_t| \ = \ \sum_{i=0}^{N-1} |a_i|^2
|s_i\rangle\langle s_i| \ ;
\end{equation}
and of the apparatus
\begin{equation}
\rho_{\cal A} \ = \mbox{Tr}_{\cal S} |\Psi_t\rangle\langle\Psi_t|
\ = \ \sum_{i=0}^{N-1}  |a_i|^2 |A_i\rangle\langle A_i| \ .
\end{equation}
\reseteqn

The correlation brought about by the interaction, Eq. (6), leads
to the loss of information about ${\cal S}$ and ${\cal A}$
individually. The entropy of each increases to
\begin{equation}
{\cal H}_{\cal S} = -\mbox{Tr}\rho_{\cal S} \log \rho_{\cal S} =
   -\sum_{i=0}^{N-1}|a_i|^2 \log |a_i|^2  =
-\mbox{Tr}\rho_{\cal A} \log \rho_{\cal A} =  {\cal H}_{\cal A} \
.
\end{equation}
As the evolution of the whole ${\cal AS}$ is unitary, the
entropies of the subsystems must be compensated by the decrease of
their mutual entropy, i.e., by the increase of their mutual
information:
\begin{equation}
{\cal I (S:A)} = {\cal H_{\cal S} + H_{\cal A} - H_{\cal SA}}
   = -2\sum_{i=0}^{N-1}|a_i|^2 \log |a_i|^2
\end{equation}
Above, $H_{\cal SA}$ is the joint entropy of ${\cal SA}$. This
quantity, Eq. (21), was introduced in the quantum context as a
measure of entanglement some time ago [3] and has been since
rediscovered and used [14].

The cost of a bit of information in terms of some other physical
quantity is an often raised question. In the context of our model
we shall inquire what is the cost of a bit transfer in terms of
action. Let us then consider a transition represented by Eq. (6).
The associated action must be no less than
\begin{equation}
I = \sum_{j=0}^{N-1} |a_j|^2 \arccos |\langle A_0|A_j\rangle|\,.
\end{equation}
When $\{ |A_j\rangle\}$ are mutually orthogonal, the action is:
\alpheqn \setcounter{saveeqn}{23}
\begin{equation}
I = \pi / 2
\end{equation}
\reseteqn in Planck ($h = 2 \pi \hbar$) units. This estimate can
be lowered by using a judiciously chosen initial $|A_0\rangle$
which is a superposition of the outcomes $| A_j\rangle$. For a
two-dimensional Hilbert space the average action can be thus
brought down to $\pi \hbar/4$ [1,3]. In general, an interaction of
the form
\begin{equation}
H_{\cal SA} = i g \sum_{k=0}^{N-1} |s_k\rangle\langle s_k|
\sum_{l=0}^{N-1} (|A_k\rangle\langle A_l| - |A_l\rangle\langle
A_k|) \
\end{equation}
saturates at the lower bound given by \alpheqn
\setcounter{saveeqn}{23} \setcounter{equation}{1}
\begin{equation}
  I = \arcsin\sqrt {1-1/N} \ .
\end{equation}
\reseteqn As the dimensionality of the Hilbert spaces increases,
the least action approaches $\pi/2$ per completely entangling
interaction. The action per bit will be less when, for a given
$N$, the transferred information is maximized, which happens when
$|a_i|^2 = 1 / N$ in Eq. (22). Then the cost of information in
Planck units is \setcounter{equation}{24}
\begin{equation}
\iota =  {I \over {\log N}} \approx {\pi \over {2 \log_2 N}}\,.
\end{equation}
The cost of information per bit decreases with increasing $N$, the
dimension of the Hilbert space of the smaller of the two
entangled systems.

This result is at the same time both enlightening and
disappointing: It shows that the cost of information transfer is
not ``fixed'' (as one might have hoped). Rather, the least total
amount of action needed for a complete entanglement is at least
asymptotically fixed as Eqs. (23) show. Consequently, the least
price per bit goes down when information is transferred
``wholesale'', i.e., when $N$ is large. Yet, this is enlightening,
as it may indicate why in the classical continuous world (where
$N$ is effectively infinite) one may be ignorant of that price
and convinced that information is free.

\section{Discord}

Mutual information can be defined either by the symmetric formula, Eq. (21),
or through an asymmetric looking equation which employs conditional entropy:
\begin{equation}
{\cal J}({\cal S:A}) = H_{\cal S} - H_{\cal S|A}
\end{equation}
Above, $ H_{\cal S | A}$ expresses the average ignorance of ${\cal S}$
remaining after the observer has found out the state of ${\cal
A}$. In classical physics, the two formulae, Eqs. (21) and (26),
are strictly identical, so that the {\it discord} between them:
\alpheqn \setcounter{saveeqn}{27}
\begin{equation}
\delta{\cal I} \ = \ {\cal I}({\cal S:A}) - {\cal J}({\cal S:A})
= 0 \
\end{equation}
always disappears [15]:
\begin{equation}
\delta{\cal I} \ = \ 0 \ .
\end{equation}
\reseteqn In quantum physics things are never this simple: To
begin with, conditional information depends on the observer
finding out about one of the subsystems, which implies a
measurement. So $H_{\cal S|A}$ must be carefully defined before
Eqs. (26) and (27) become meaningful.

Conditional entropy is non-trivial in the quantum context
because, in general, in order to find out  $ H_{\cal S | A}$  one
must choose a set of projection operators $\Pi_j$ and define a
conditional density matrix given by the outcome corresponding to
$\Pi_j$ through
\begin{equation}
\tilde \rho_{{\cal S}|\Pi_j} = \mbox{Tr}_{\cal A} \Pi_j \rho_{\cal
SA}\,,
\end{equation}
where in the simplest case $\Pi_j = |C_j\rangle\langle C_j|$ is a
projection operator onto a pure state of the apparatus. A
normalized $\rho_{{\cal S}|\Pi_j}$ can be obtained by using the
probability of the outcome
\begin{equation}
p_j = \mbox{Tr} \tilde \rho_{{\cal S}|\Pi_j} \ ;
\end{equation}
\begin{equation}
\rho_{{\cal S}|\Pi_j} =p_j^{-1}\tilde \rho_{{\cal S}|\Pi_j} \ .
\end{equation}

The conditional density matrix $\rho_{{\cal S}|\Pi_j} $ represents
the description of the system ${\cal S}$ available to the
observer who knows that the apparatus ${\cal A}$ is in a subspace
defined by $\Pi_j$. For a pure initial state and an exhaustive
measurement the conditional density matrix will also be pure. We
shall however consider a broader range of possibilities,
including joint density matrices which undergo a decoherence
process, so that
\begin{equation}
\rho^P_{\cal SA} = \sum_{i,j} \alpha_i \alpha_j^*
|s_i\rangle\langle s_j||A_i\rangle\langle A_j| \Longrightarrow
\sum_i |\alpha_i|^2 |s_i\rangle\langle s_i||A_i\rangle\langle A_i|
= \rho_{\cal SA}^D\,.
\end{equation}
This transition is accompanied by an increase in entropy
\begin{eqnarray*}
\Delta H(\rho_{\cal SA})=H(\rho_{\cal SA}^D) - H({\rho^P_{\cal
SA}})
\end{eqnarray*}
and by the simultaneous disappearance of the ambiguity in what
was measured [1-3]. Now, $\rho_{{\cal S}|\Pi_j}$ is no longer
pure, unless $\Pi_j = |A_j\rangle\langle A_j|$. That is, a
measurement of the apparatus in bases other than the pointer
basis will leave an observer with varying degrees of ignorance
about the state of the system. More general cases when the
density matrix is neither a pure pre-decoherence projection
operator $ \rho^P_{\cal SA} $ to the left of the arrow in Eq.
(31) nor a completely decohered $\rho_{\cal SA}^D$ state on the
right are possible and typical.

To define discord $\delta {\cal I}$ we finalize our definition of
$H_{\cal S|A}$:
\begin{equation}
  H_{\cal S|A} = \sum_i p_{|C_i\rangle} H(\rho^D_{{\cal S}|C_i\rangle}) \ .
\end{equation}
Above, we have used an obvious notation for the density matrix
conditioned upon pure states $\{ |C_j\rangle \}$. We emphasize
again that the conditional entropy depends on $\rho_{\cal SA}$,
but also on the choice of the observable measured on ${\cal A}$.
In classical physics all observables commute, so there is no such
dependence. Thus, non-commutation of observables in quantum
theory is the ultimate source of the information - theoretic
discord.

The obvious use for the discord is to employ it as a measure of
how non-classical the underlying correlation of two quantum
systems is. In particular, when there exists a set of states in
one of the two systems in which the discord disappears, the state
represented by $\rho_{\cal SA}$ admits a classical interpretation
of probabilities in that special basis. Moreover, unless the
discord disappears for trivial reasons (which would happen in the
absence of correlation, i.e., when $\rho_{\cal SA} = \rho_{\cal
S} \rho_{\cal A}$), the basis which minimizes the discord can be
regarded as ``the most classical". For $\delta{\cal I} = 0$ the
states of such preferred basis and their corresponding
eigenvalues can be treated as effectively classical [11].

The vanishing discord is a stronger condition than the absence of
entanglement. In effect, $\delta{\cal I} = 0$ implies existence
of the eigenstates of $\rho_{\cal SA}$ which are products of the
states of ${\cal S}$ and of ${\cal A}$. An instructive example of
this situation arises as a result of decoherence: When the
off-diagonal terms of $\rho_{\cal SA}$ disappear in a manner
illustrated by Eq. (31), the discord disappears as well.

\section{Environment as a witness: Redundancy ratio}

The discussion of decoherence to date tends to focus on the
effect of the environment on the system or on the apparatus. The
destruction of quantum coherence and the emergence of preferred
pointer observables whose eigenvalues are associated with the
decoherence-free pointer subspaces was the focus of the
investigation.

Here we shall break with this tradition. According to the theory
of decoherence, the environment is monitoring the system.
Therefore, its state must contain a record of the system. It is
of obvious interest to analyze the nature and the role of this
record. To this end, we shall use mutual information introduced
before defining the {\it redundancy ratio}
\begin{equation}
  {\cal R}_{{\cal I}{\left(\otimes {\cal H}_{{\cal E}_k} \right)}}
= \left (\sum_{k} \ {\cal I(S:E}_k)\right)/ {\cal H(S)} \ \ .
\end{equation}
Above, we imagined a setting where the system is decohering due to
the interaction with the environment which is composed of many
subsystems ${\cal E}_k$. ${\cal R}_{{\cal I}{\left (\{\otimes
{\cal H}_{{\cal E}_k}\} \right)}} $ is a measure of how many
times -- how redundantly -- the information about the system has
been inscribed in the environment. An essentially identical
formula can be introduced using the asymmetric ${\cal J}$, Eq.
(26). It is easy to establish that the discord is always
non-negative and, hence, that
\begin{equation}
{\cal R}_{{\cal J}_{MAX}} \leq {\cal R}_{{\cal I}_{MAX}}  \ .
\end{equation}
The subscript indicating maximization may refer to two distinct
procedures: ${\cal R_J}$ will obviously depend on the manner in
which subsystems of the environment are measured. In fact, it is
convenient to use
\begin{equation}
{\cal J}_k = {\cal J}({\cal S : E}_k) =
%{\cal H}({\cal S}) - {\cal H} ({\cal S}|{\cal E}_k) =
{\cal H}({\cal E}_k) - {\cal H} ({\cal E}_k|{\cal S}) \ \ ,
\end{equation}
to define a basis-dependent
\begin{equation}
{\cal R}_{\cal J}(\{|s\rangle\}) = {\cal R}_{\cal J} (\otimes
{\cal H_E}_k)
\end{equation}
in a manner analogous to Eq. (33). Maximizing ${\cal R}_{\cal
J}(\{|s\rangle\}) $ with respect to the choice of the choice of
states $\{ |s\rangle \}$ [11] is an obvious ``counterpoint" to the
predictability sieve [16-19], the strategy which seeks states
that entangle the least with the environment.

There is one more maximization procedure which may and should be
considered: The environment can be partitioned differently -- for
example, it may turn out that more information about the system
can be extracted by measuring, say, the photon environment in
some collective fashion (homodyne?) instead of directly counting
  the environment photons. It is clear that for such
optimization to be physically significant, it should respect to
some degree the natural structure of the environment.

In addition to the redundancy ratio one can define the rate at
which the redundancy ratio increases. The redundancy rate is
defined as
\begin{equation}
\dot{{\cal R}} = \frac{d}{dt} {\cal R}\,.
\end{equation}
Either the basis dependent or the basis-independent versions of
$\dot {\cal R}$ may be of interest. The physical significance of
the redundancy ratio rate is clear: It shows how quickly the
information about the system spreads throughout the environment.
It is, in effect, a measure of the rate of increase of the
effective number of the environment subsystems which have recorded
the state of the system ${\cal S}$.

It is worth noting that either ${\cal R_{\cal I}}$ or ${\cal
R_{\cal J}}$ can keep on increasing after the density matrix of
the system has lost its off-diagonal terms in the pointer basis
and after it can be therefore considered completely decohered.
Indeed, direct interaction between the system and the environment
is not needed for either ${\cal R}$ to change. For example,
information about the system inscribed in the primary environment
may be communicated to a secondary, tertiary, and more remote
environment (which need not interact with the system at all).

It is natural to define objectivity and, therefore, classicality
with the help of ${\cal R}$. In the limit ${\cal R_J} \rightarrow
\infty$ the information about the preferred states of the system
is spread so widely that it can be acquired by many observers
simultaneously [11]. Moreover, it already exists in multiple copies,
so it can be safely cloned in spite of the no-cloning theorem [20].
Thus, a state of the system redundantly recorded in the environment has
all the symptoms of ``objective existence''. In particular, such
well-advertised states can be found out without being disturbed
by approximately ${\cal R}$ observers acting independently
[11,21] (each simply measuring the state of $\sim 1/{\cal R_J}$
fraction of the environment).

\section{Summary and Conclusions}

Information theory offers a useful perspective on the measurement
process, on decoherence and, above all, on the definitions of
classicality. Discord can be used to characterize the nature of
quantum correlations and to distinguish the ones that are
classical. The redundancy ratio is a powerful measure of the
classicality of states: While a vanishing discord is a necessary
condition for classicality of correlations, the redundancy ratio
is a direct measure of objective existence of quantum states.
Objectivity can be defined operationally as the ability
to find what the state is without disturbing it [11,21].
Objective existence of quantum states would make them effectively
classical, and was the ultimate goal of the interpretation of
quantum theory. We have established it here by investigating
einselection from the point
of view of information theory and by shifting focus from the
system to the environment which is monitoring the system.

These advances clarify some of the interpretational issues which are now
a century old. The relation between the epistemological and ontological
significance of the quantum state vectors is now apparent: Objective existence
of the quantum states is a direct consequence of the redundant records
permeating the environment. Epistemology begets ontology!

This research was supported in part by NSA.


\begin{thebibliography}{99}

\bibitem{1} Zurek, W. H., Phys. Rev. D {\bf 24} (1981) 1516-1524

\bibitem{2} Zurek, W. H., Phys. Rev. D {\bf 26} (1982) 1862-1880

\bibitem{3} Zurek, W. H., Information transfer in quantum
measurements, pp. 87-116 in {\it Quantum Optics, Experimental
Gravity, and the Measurement Theory}, P. Meystre and M. O.
Scully, eds. (Plenum, New York, 1983)

\bibitem{4} Schumacher, B., Westmoreland, M., and Wootters, W. K.,
Phys. Rev. Lett. {\bf 76} (1996) 3452-3455; Hall, M. J. H., and
O'Rourke, M. J., Quant. Opt. {\bf 5} (1993) 161; Halliwell, J. J.,
Phys. Rev. D {\bf 60} (1999) 105031

\bibitem{5} Joos, E., and Zeh, H. D., Zeits. Phys. B {\bf 59}
(1985) 229

\bibitem{6} Caldeira, A. O., and Leggett, A. J., Physica {\bf
121A} (1983) 587-616;  Phys. Rev. A {\bf 31} (1985) 1059; Haake,
F., and Reibold, R., Phys. Rev. A {\bf 32} (1985) 2462; Grabert,
H., Schramm, P., and Ingold., G.-L., Phys. Rev. {\bf 168} (1988)
115-207

\bibitem{7} Zurek, W. H., Reduction of the Wavepacket: How long
does it take? presented at a NATO ASI {\it Frontiers of
Nonequilibrium Statistical Mechanics} Santa Fe, June 1984, pp.
145-149 in the proceedings, G. T. Moore and M. O. Scully, eds.
(Plenum, New York, 1986); Haake, F., and Walls, D. F., in {\it
Quantum Optics IV}, J. D. Harvey and D. F. Walls, eds. (Springer,
Berlin, 1986). Unruh, W. G., and Zurek, W. H., Phys. Rev. D {\bf
40} (1989) 1071-1094


\bibitem{8} Hu, B. L., Paz, J. P., and Zhang, Y., Phys. Rev. D {\bf 45}
(1992) 2843-2861; Zurek, W. H., Habib, S., and Paz, J. P., Phys.
Rev. Lett {\bf 70} (1993) 1187-1190; Anglin, J. R., and Zurek, W.
H., Phys Rev. D {\bf 53} (1996) 7327-7335; Anglin, J. R., Paz, J.
P., and Zurek, W. H., Phys. Rev. A {\bf 53} (1997) 4041

\bibitem{9} Brune, M., Hagley, E., Dreyer, J., Ma\^itre, X., Maali, A.,
Wunderlich, C., Raimond, J-M., and Haroche, S., Phys. Rev. Lett.
{\bf 77} (1996) 4887-4890; Myatt {\it et al.}, Nature {\bf 403}
(2000) 269

\bibitem{10} Giulini, D., Joos, E., Kiefer, C., Kupsch, J.,
Stamatescu, I.-O., and
Zeh, H. D., {\it Decoherence and the Appearance of a Classical World in Quantum
Theory}, (Springer, Berlin, 1996);
Blanchard, Ph., Giulini, D., Joos, E., Kiefer, C.,
and Stamatescu, I.-O., eds, {\it Decoherence: Theoretical, Experimental,
and Conceptual Problems}, (Springer, Berlin, 2000);
Paz, J. P., and Zurek, W. H., in {\it Les Houches Lectures}, in press (2000)

\bibitem{11} Zurek, W. H., Rev. Mod. Phys., submitted (2000)

\bibitem{12} Ivanovic I. D.,  J. Phys. A {\bf 14} (1981) 3241-3245;
J. Math. Phys. {\bf 24} (1983) 1199-1205

\bibitem{13} Wootters, W. K., and Fields, B. D., Ann. Phys. {\bf 191}
(1989) 363

\bibitem{14} Barnett, S. M., and Phoenix, S. J. D., Phys. Rev. A {\bf 40}
(1989) 2404-2409

\bibitem{15} Cover, T. M., and Thomas, J. A., {\it Elements of
Information Theory},
(Wiley, New York, 1991)

\bibitem{16} Zurek, W. H., Progr. Theor. Phys. {\bf 89} (1993) 281-302

\bibitem{17} Zurek, W. H., Habib, S., and Paz, J. P., Phys. Rev.
Lett {\bf 70} (1993) 1187-1190

\bibitem{18} Tegmark, M., and
Shapiro, H. S., Phys. Rev. E {\bf 50} (1994) 2538-2547

\bibitem{19} Gallis, M. R., Phys. Rev. A {\bf  53} (196) 655-660;
Paraoanu, Gh.-S., and Scutaru, H., Phys. Lett. A {\bf  238} (1998)
219-222

\bibitem{20} Wootters, W. K., and Zurek, W. H., Nature {\bf
299} (1982) 802-802; Dieks, D.,  Phys. Lett. {\bf 92} (1982)
271-272

\bibitem{21} Zurek, W. H., Phil. Trans. Roy. Soc. Lond. {\bf
356} (1998) 1793-1821

\end{thebibliography}
\end{document}